\documentclass[12pt]{article}

\usepackage{epsf}
\usepackage{latexsym,euscript}
\usepackage[dvips]{graphicx}
\usepackage{amsmath}
\usepackage{amsfonts}
\usepackage{amssymb, epsfig}
\usepackage{cite}
\usepackage{subcaption}
\usepackage[]{units}
\usepackage{mwe}
\usepackage{multirow}
\usepackage{verbatim}
\usepackage{geometry}
\usepackage{color,soul}

\usepackage{longtable}
\usepackage{booktabs}
\usepackage{float}
\usepackage{lscape}
\usepackage{subcaption}
\usepackage{cleveref}

\DeclareMathOperator{\arcsinh}{arcsinh}
\DeclareMathOperator{\Tr}{Tr}

\renewcommand{\Re}{\operatorname{Re}}
\renewcommand{\Im}{\operatorname{Im}}

\DeclareMathOperator{\TR}{Tr}
\DeclareMathOperator{\RE}{Re}
\DeclareMathOperator{\IM}{Im}

\begin{document}

\begin{center}
  {\Large \bf Dual simulation of a Polyakov loop model at finite baryon
  	density: \\ \vspace{0.1cm} correlations and screening masses}
\end{center}

\vskip 0.3cm
\centerline{O.~Borisenko$^{1,2\dagger}$, V.~Chelnokov$^{3*}$,
	E.~Mendicelli$^{4\ddagger}$, A.~Papa$^{1,5\P}$}

\vskip 0.6cm

\centerline{${}^1$ \sl Istituto Nazionale di Fisica Nucleare,
	Gruppo collegato di Cosenza,}
\centerline{\sl I-87036 Arcavacata di Rende, Cosenza, Italy}

\vskip 0.2cm

\centerline{${}^2$ \sl Bogolyubov Institute for Theoretical Physics,}
\centerline{\sl National Academy of Sciences of Ukraine,}
\centerline{\sl 03143 Kyiv, Ukraine}

\vskip 0.2cm

\centerline{${}^3$ \sl Institut f\"ur Theoretische Physik, Goethe-Universit\"at
	Frankfurt,}
\centerline{\sl Max-von-Laue-Str. 1, 60438 Frankfurt am Main, Germany}

\vskip 0.2cm

\centerline{${}^4$ \sl Department of Physics and Astronomy, York University,}
\centerline{\sl Toronto, ON, M3J 1P3, Canada}

\vskip 0.2cm

\centerline{${}^5$ \sl Dipartimento di Fisica, Universit\`a della
	Calabria,}
\centerline{\sl I-87036 Arcavacata di Rende, Cosenza, Italy}

\vskip 0.6cm

\begin{abstract} 
Computations of screening masses in finite-temperature QCD at finite density are plagued by the sign problem and have  been
performed so far with an imaginary chemical potential. Here, we use a dual formulation of a Polyakov-loop model which allows
the determination of screening masses at real baryon chemical potential. This is a second paper in a series devoted to a
detailed study of dual Polyakov-loop models at finite density. While the first paper was mainly devoted to establishing the
phase diagram of the model, here we compute correlation functions of the Polyakov loops and the second-moment correlation 
length at non-zero chemical potential. This enables us to evaluate numerically the screening masses from correlations of the
real and imaginary parts of the Polyakov loops. We also compute these masses in the mean-field approximation and compare with numerical
results. In addition, we provide a quantitative improvement of the general phase diagram presented in the first paper.  
\end{abstract}

\vfill
\hrule
\vspace{0.3cm}

{\it e-mail addresses}:
$^\dagger$oleg@bitp.kiev.ua, \  $^*$chelnokov@itp.uni-frankfurt.de, \\
$^{\ddagger}$emanuelemendicelli@hotmail.it, \ \ $^{\P}$alessandro.papa@fis.unical.it

\newpage

\section{Introduction}

Understanding the properties of strongly interacting matter at finite
temperatures and densities is exceedingly important in many areas of
research, including cosmology, astrophysics of compact objects and
phenomenology of heavy-ion collisions. Numerical investigations on a
Euclidean space-time lattice, based on Monte-Carlo methods, are hindered by
the notorious sign problem: in presence of a non-zero baryon chemical potential
the Boltzmann weight turns complex and cannot be used to sample thermal ensemble
configurations. Many approaches have been designed in the last decades to
circumvent or mitigate this problem, such as Taylor expansion around
zero baryon chemical potential, reweighting to small baryon chemical potential,
simulations at imaginary potential, complex Langevin simulations and others 
(see, {\it e.g.},~\cite{philipsen19,Schmidt:2006us,Forcrand_rev_10,Seiler_rev_17}). 

Recent times have witnessed the development of radically new approaches that 
aim to fully solve the sign problem by expressing the original
partition function and observables in terms of different, usually
integer-valued, degrees of freedom, so to make the resulting Boltzmann weight
positive definite. These formulations are conventionally called ``dual'',
even though sometimes different names are used, such as ``flux line
representation''~\cite{Gattringer_rev_16}.

While several strategies have been developed to get dual formulations for
non-Abelian lattice models with fermions~\cite{su3_abc,un_dual,Unger_19},
a dual formulation with a positive Boltzmann weight is not yet available
for full QCD. In some limiting regimes, however, it has been possible to
construct formulations either positive or such that the sign problem is
less severe. This is the case of the strong-coupling limit of QCD, where the
$SU(N)$ lattice gauge theory (LGT) can be mapped onto a monomer-dimer and
closed baryon loop model~\cite{Karsch_89} (see also Ref.~\cite{Unger_20} and
references therein). Other cases include many effective Polyakov-loop
models which can be derived from the full lattice QCD in some specific
limits. Examples of such dual models with positive Boltzmann weight are known 
in the wider context of the $SU(N)$~\cite{Gattringer11,Gattringer12,Philipsen12,Borisenko20}
and $U(N)$ groups~\cite{un_dual}.

In this paper we focus on a 3-dimensional effective Polyakov-loop model,
describing the $(3+1)$-dimensional $SU(N)$ LGT with one flavor of
staggered fermions at finite baryon density. It is defined on a $3$-dimensional
hypercubic lattice $\Lambda = L^3$, with $L$ the linear extension and
a unit lattice spacing; $\vec{x}\equiv x=(x_1,x_2,x_3)$, $x_i\in [0,L-1]$
denote the sites of the lattice, $l=(\vec{x},\nu)$ is the lattice link in the
direction $\nu$; $e_{\nu}$ is a unit vector in the direction $\nu$ and $N_t$
is the lattice size in the temporal direction of the underlying (3+1)-dimensional
LGT; periodic boundary conditions are imposed in all directions. 
The general form of the partition function of the model reads~\cite{Caselle97,Philipsen14} 
\begin{eqnarray}
Z_{\Lambda}(\beta,\bar{m},\bar{\mu};N)  &\equiv&  Z \ = \ \int \prod_x dU(x)
\exp \left [ \beta \sum_{x,\nu} \ {\rm {Re}}{\rm {Tr}}U(x){\rm {Tr}}U^{\dagger}(x+e_\nu) \right ] \nonumber  \\ 
&\times& \ \prod_x \ A(\bar{m}) \ {\rm det} \left [ 1 + h_+ U(x) \right ] \ {\rm det} \left [ 1 + h_- U^{\dagger}(x) \right ] \ . 
\label{sunpf}
\end{eqnarray}
In this model the matrices $U(x)$ play the role of Polyakov loops, the only
gauge-invariant operators surviving the integration over spatial gauge fields
and over quarks. ${\rm Tr}U$ denotes the fundamental character of $SU(N)$. 
Integration in~(\ref{sunpf}) is performed with respect to the Haar measure on
$SU(N)$. The effective coupling constant $\beta$ is a complicated function of the original gauge coupling (its precise form is not important here). 
The constants $A(\bar{m})$ and $h_{\pm}$ are given by 
\begin{equation}
A(\bar{m}) \ = \ e^{N N_t \arcsinh a \bar{m} } \ \approx \ e^{\frac{N \bar{m}}{T}} \ , \ h_{\pm} \ = \ e^{-(\arcsinh a \bar{m} \mp a \bar{\mu}) N_t} \ \approx \ e^{-\frac{\bar{m} \mp \bar{\mu}}{T}} \ . 
\label{hpm_stag}
\end{equation}
The pure-gauge part of the effective action is invariant only
under global discrete transformations $U(x)\to Z U(x)$, $Z\in Z(N)$. 
This is the global $Z(N)$ symmetry. The quark contribution violates this
symmetry explicitly. Another important feature of the Boltzmann weight is that it becomes complex in the presence of a chemical potential, as follows from~(\ref{sunpf}). Therefore, the model cannot be studied by direct Monte-Carlo methods if $\mu$ is non-zero. 

The model defined in Eq.~(\ref{sunpf}) has been investigated in the large-$N$
limit of $U(N)$ at non-zero $\mu$ in~\cite{Christensen:2012km}; it was found
that the dependence on the chemical potential drops out from the free energy.
Moreover, a third-order phase transition has been reported in this limit. 
The large-$N$ limit of the $SU(N)$ case was studied in~\cite{LargeN_PLmodel,LargeN_PLmodel_corr}.
In Refs.~\cite{Greensite12,Greensite14} a mean-field approximation was used
to study the $SU(3)$ model. 

Dual formulations with positive Boltzmann weight of the model defined in
Eq.~(\ref{sunpf}) have been constructed in Refs.~\cite{Gattringer11,un_dual},
thus making Monte-Carlo simulations feasible~\cite{Philipsen12,Gattringer12,Delgado12}.
Mean-field and Monte-Carlo studies turned out to be in quantitative agreement for
the energy density and expectation values of the Polyakov loops. In some regions
of the parameters $\beta$, $h$ and $\mu$ the phase diagram of the model has been
revealed, but long-distance correlations and, hence the screening masses, have
not been computed so far (see, however,~\cite{Borisenko19}). 

In a previous paper~\cite{Borisenko:2020cjx}, we have studied an alternative
dual form of the partition function~(\ref{sunpf}), originally derived
in~\cite{Borisenko20} (its explicit expression is given below).
By a finite-size scaling of the magnetization susceptibility, we have studied
in great details the phase diagram of the model, identifying
three regions in the parameter space $(\beta, h, \mu)$ according to the type
of the critical behavior: first or second order phase transition, or crossover. 
We have computed expectation values of Polyakov loop, baryon density and
quark condensate, finding good agreement with mean-field predictions
both at zero and non-zero $\mu$. In general, we have found that the overall
qualitative picture of the phase diagram and the behavior of all observables
fully agree with Refs.~\cite{Philipsen12,Delgado12}. We have also observed
that, in general, all the observables considered in that work are sensitive
to the chemical potential, exhibiting a less steep variation across transition
when the coupling $\beta$ (which corresponds to the temperature in the
underlying QCD theory) is increased. Qualitatively, the effect of increasing
$\mu$ has shown to play the same role of the reduction of the quark mass.

In the present paper we exploit the same dual form of the effective 
Polyakov-loop model adopted in~\cite{Borisenko:2020cjx} to study several correlations of the Polyakov loops and to extract screening (electric and magnetic) masses at finite chemical potential. We consider also the determination of the
second-moment correlation length. To the best of our knowledge, this kind of
analysis has never been done so far in the framework of dual formulations
of $SU(N)$ effective Polyakov-loop models. It is important not only to size
the effect of a non-zero baryon chemical potential on screening effects in
the deconfined phase, but also for the investigation of the elusive
oscillating phase at finite
density~\cite{Ogilvie10,Ogilvie16,oscillating_phase}, which is ultimately
connected to the complex spectrum of the theory. 

For a general review of the screening masses in lattice QCD at finite temperature we refer to~\cite{Bazavov:2020teh}. Simulations at imaginary chemical potential with two flavors of Wilson fermions and with $(2+1)$ flavor of staggered fermions have been performed in Ref.~\cite{Nagata13} and Ref.~\cite{bonati2018}, correspondingly. 

We describe now the dual form of the partition
function~(\ref{sunpf}). This dual representation will be used 
in the next Sections for numerical simulations of the model. All details of the
derivation can be found in~\cite{Borisenko20}. 
In the case of one flavor of staggered fermions the partition
function~(\ref{sunpf}) can be presented, after an exact integration over
Polyakov loops, as  
\begin{equation} 
	\label{PF_statdet_3}
	Z = \sum_{\{ r(l) \} = -\infty}^{\infty} \ \sum_{\{ s(l) \}= 0}^{\infty} \ 
	\prod_l  \frac{\left ( \frac{\beta}{2} \right )^{| r(l) | + 2s(l)}}{(s(l)+| r(l) |)!
		s(l)!} \ 
	\prod_x A(\bar{m}) R_N(n(x),p(x))  \ , 
\end{equation}
\begin{eqnarray}
	\label{nx_stat}
	&&n(x) =  \sum_{i=1}^{2d}  \left ( s(l_i) + \frac{1}{2} | r(l_i) | \right )
	+ \frac{1}{2} \sum_{\nu=1}^{d} \left ( r_{\nu}(x) - r_{\nu}(x-e_{\nu}) \right )
	\ ,   \\ 
	\label{px_stat}
	&&p(x) =  \sum_{i=1}^{2d}  \left ( s(l_i) + \frac{1}{2} | r(l_i) | \right )
	- \frac{1}{2} \sum_{\nu=1}^{d} \left ( r_{\nu}(x) - r_{\nu}(x-e_{\nu}) \right )
	\ ,  
\end{eqnarray} 
where $l_i, i=1,...,2d$ are $2d$ links attached to a site $x$ and 
\begin{eqnarray} 
	R_N(n,p) = \sum_{q=-\infty}^{\infty} \sum_{k,l=0}^N \sum_{\sigma \vdash n+k}
	\ \delta_{n+k,p+l+qN} \ 
	d(\sigma/1^k) d(\sigma+q^N/1^l) \ h_+^k h_-^l \ . 
	\label{Rint_Nf1}
\end{eqnarray}
The sum over $\sigma$ runs over all partitions of $n+k$, and
$d\left ( \sigma/1^m  \right )$ is the dimension of a skew representation
defined by a corresponding skew Young diagram,
$\sigma+q^N=(\sigma_1+q,\ldots,\sigma_N+q)$
(for more details we refer the reader to Ref.~\cite{Borisenko20}).
Equation~({\ref{PF_statdet_3}}) is valid for all $SU(N)$ groups and in any
dimension. 
Clearly, all factors entering the Boltzmann weight of~(\ref{PF_statdet_3}) are
positive. 
Hence, this representation is suitable for numerical simulations.
The Kronecker delta-function in expression~(\ref{Rint_Nf1}) represents the
$N$-ality constraint on the admissible configurations of the integer-valued
variables $s(l)$ and $r(l)$. 
This constraint can be exactly resolved only in the pure gauge model when
$h_{\pm}=0$. 
In this case the dual representation~(\ref{PF_statdet_3}) has been already
tested by us on the example  of a 2-dimensional $SU(3)$ model, where we studied
correlation functions and three-quark potential~\cite{Borisenko19}. 

In the following Sections we study the dual representation~(\ref{PF_statdet_3})
via Monte-Carlo simulations for the 3-dimensional $SU(3)$ model. In this
case the function $R_N(n,p;h_{\pm})$ takes the form 
\begin{eqnarray} 
	\label{Rint_N3_Nf1}
	&&R_3(n,p) = Q_3(n+1,p) \left ( h_+ + h_-^2 + h_+ h_-^3 + h_+^3 h_-^2  \right ) \\
	&+&Q_3(n,p) \left ( 1+ h_+^3 + h_-^3 + h_+^3 h_-^3  \right )  + 
	Q_3(n,p+1) \left ( h_- + h_+^2 + h_+^3 h_- + h_+^2 h_-^3  \right )   \nonumber \\ 
	&+&Q_3(n+1,p+1) \left ( h_+ h_- + h_+^2 h_-^2  \right ) + 
	Q_3(n+2,p) h_+ h_-^2  + Q_3(n,p+2) h_+^2 h_- \ .   \nonumber  
\end{eqnarray}
The function $Q_3(n,p)$ is the result of the group integration and is given
by~\cite{weingarten_sun}
\begin{equation} 
	Q_N(n,p) \ = \  
	\sum_{\lambda \vdash {\rm min}(n,p)} \ d(\lambda) \ d(\lambda + |q|^N) \ , 
	\label{QSUN}
\end{equation}
where $d(\lambda)$ is the dimension of the permutation group $S_r$ in
the representation $\lambda$, $q = (p - n) / N$ (when $q$ is not an integer,
$Q_N(n,p) = 0$).

Important is the fact that {\em both} local observables and long-distance
quantities can be computed with the help of this dual representation. Explicit
expressions for the correlation functions of the Polyakov loops 
will be given in Section~3.

The paper is organized as follows. In the next Section we present a mean-field study of the correlations for $SU(3)$ Polyakov-loop model and calculate the screening masses analytically. In Section~3 we give the definitions of several Polyakov-loop correlators both in the standard and in the dual formulations; we define also in this context the second-moment correlation length. Then, we present our numerical Monte-Carlo results for Polyakov-loop correlation functions,
screening masses and second-moment correlation length. In Section~4 we summarize our main results and outline the future work. In Appendix we present improved phase diagram of the Polyakov-loop model.

\section{Analytic study of the correlations}

First of all, we would like to derive some analytic predictions for the behavior of the  correlations and screening masses. Here we calculate the partition and correlation functions of $SU(3)$ model following the approach of Refs.~\cite{LargeN_PLmodel,LargeN_PLmodel_corr} developed for the large-$N$ limit. 
All results presented in this and next Sections are given in terms
of dimensionless quantities $m=\bar{m}/T$ and $\mu=\bar{\mu}/T$.

Consider the following change of variables: 
\begin{eqnarray}
{\rm Re}{\rm Tr}U(x) &=& 3 \rho(x) \cos\omega(x)   \  ,  \nonumber  \\ 
{\rm Im}{\rm Tr}U^{\dagger}(x) &=& 3 \rho(x) \sin\omega(x) \ . 
\label{change_var}
\end{eqnarray}
The partition function can then be rewritten as 
\begin{equation}
Z  \ = \ \int \prod_x \rho(x) d\rho(x) \ \frac{d\omega(x)}{2\pi} \ 
\exp \left [  S(\rho(x),\omega(x))  \right ] . 
\label{PF_2}
\end{equation}
The full action in new variables takes the form 
\begin{equation}
S(\rho(x),\omega(x)) \ = \ S_g + \sum_x \left ( \ln \sqrt{J(x)} + \ln 2 + 
\frac{3}{2}\ln 3 + \ln Q(x) \right ) \ , 
\label{action_eff}
\end{equation}
where $S_g$ is the gauge part of the action 
\begin{equation}
S_g \ = \ 9\beta \sum_{x,n} \rho(x) \rho(x+e_n) \cos(\omega(x) - \omega(x+e_n)) \ . 
\label{S_gauge}
\end{equation}
$J(x)$ is the Jacobian of the transformation~\cite{Wipf_06}
\begin{equation}
J(x) = 1 - \frac{3}{4} \ \left ( 1+\rho^2(x) \right )^2 +  
2 \rho^3(x) \cos 3\omega(x) \ 
\label{Jacobian_1}
\end{equation}
and $Q(x)$ represents the quark contribution 
\begin{eqnarray}
Q(x) &=& \cosh 3m + \cosh 3\mu + 6 \cosh 2m \ \rho(x) \cos(\omega(x) - i\mu) + 
9 \cosh m \ \rho^2(x) \nonumber  \\ 
&+& 6 \cosh m \ \rho(x) \cos(\omega(x)+2 i\mu) + 
18 \rho^2(x) \cos(2 \omega(x)+i\mu)  \ . 
\label{Q_quark_full}
\end{eqnarray}
The integration in~(\ref{PF_2}) is performed over the domain where $J(x) > 0$. 

Firstly, we look for constant, translation invariant solutions of the saddle-point equations 
\begin{eqnarray}
\label{saddle-eqns_1}
18 d \beta \rho + \frac{1}{2 J} \ \frac{\partial}{\partial \rho} \ J + 
\frac{1}{Q} \ \frac{\partial}{\partial \rho} \ Q = 0 \ ,  \\ 
\frac{1}{2 J} \ \frac{\partial}{\partial \omega} \ J + 
\frac{1}{Q} \ \frac{\partial}{\partial \omega} \ Q = 0 \ .
\label{saddle-eqns_2}
\end{eqnarray}
Secondly, we expand the full action around the saddle points. If $\rho_0,\omega_0$ are the saddle points, we make the substitution $\rho(x)\to\rho_0+\frac{1}{3}\rho(x)$, $\omega(x)\to\omega_0 + \frac{1}{3}\omega(x)$ and expand in powers of fluctuations. The action becomes 
\begin{eqnarray}
\label{action_expansion}
&&S(\rho(x),\omega(x)) = S_{cl}(\rho_0,\omega_0) + \beta \sum_{x,n} \rho(x)\rho(x+e_n) \\ 
&& - \frac{1}{2}\ \beta \rho_0^2 \sum_{x,n} \left ( \omega(x) - 
\omega(x+e_n)  \right )^2 - \sum_x \left ( b_1 \rho^2(x) + b_2 \omega^2(x) 
- b_3 \rho(x) \omega(x)  \right ) \ . 
\nonumber  
\end{eqnarray}
Here, $S_{cl}$ is the classical action and coefficients $b_i$ are functions of $\rho_0,\omega_0$. These coefficients are very lengthy and cumbersome, so we choose not to give here their precise expressions. Important is that $b_1$ and $b_2$ are always real and non-vanishing, while $b_3$ is purely imaginary and non-zero only when the chemical potential is non-vanishing. Finally, we calculate the correlation functions by integrating over quadratic fluctuations. For the arbitrary normalized correlations of the fundamental characters of the Polyakov loops 
\begin{eqnarray}
\Gamma(\eta,\bar{\eta}) = \langle \ \prod_x \rho^{\eta(x)+\bar{\eta}(x)}(x) \ 
e^{i\omega(x) (\eta(x)-\bar{\eta}(x))}   \ \rangle \;,
\label{correlator_def}
\end{eqnarray}
one obtains 
\begin{eqnarray}
\label{correlator_res}
\Gamma(\eta,\bar{\eta}) =  \prod_x \rho_0^{\eta(x)+\bar{\eta}(x)} 
e^{i\omega_0 (\eta(x)-\bar{\eta}(x))} 
\exp\left [ \frac{1}{4 N^2 \beta \rho_0^2}  \sum_{x,x^{\prime}} 
\left (  A_1(x,x^{\prime}) + A_2(x,x^{\prime}) \right ) \right ]  ,  
\end{eqnarray}
\begin{eqnarray}
\label{A1_notation}
A_1(x,x^{\prime}) &=& \!\!\frac{1}{\sqrt{C_1 C_2}} \left ( C_1 \eta(x)\eta(x^{\prime}) 
+ C_2 \bar{\eta}(x)\bar{\eta}(x^{\prime})  \right )
\left ( G_{x,x^{\prime}}(m_1) - G_{x,x^{\prime}}(m_2) \right )\!, \\
\label{A2_notation}
A_2(x,x^{\prime}) &=& 2 \eta(x) \bar{\eta}(x^{\prime}) \ 
\left ( G_{x,x^{\prime}}(m_1) + G_{x,x^{\prime}}(m_2) \right )  \ .  
\end{eqnarray}
We have used the following notation for the Green function: 
\begin{equation}
G_{x,x^{\prime}}(m) = \frac{1}{L^d} \ \sum_{k_n=0}^{L-1} \ 
\frac{e^{\frac{2\pi i}{L} \sum_n^d k_n (x_n-x_n^{\prime})}}{m + f(k)} \ , \ f(k) = d-\sum_{n=1}^d\cos\frac{2\pi}{L}k_n \ . 
\label{Green_func}
\end{equation}
Constants $C_{1,2}$ define the screening masses and are given below. 
This general result allows one to compute any invariant or non-invariant observable by choosing the appropriate values of the sources $\eta(x)$ and $\bar{\eta}(x)$. 
{\it E.g.}, the magnetization $M=\langle \rho(x) e^{i\omega(x)} \rangle$ and its complex conjugate become 
\begin{eqnarray}
\label{magnetization}
M = \rho_0 e^{i\omega_0} \ \exp\left [ \frac{C_1}{4 N^2 \beta \rho_0^2 \sqrt{C_1 C_2}} \ \left (   G_0(m_1) - G_0(m_2) \right ) \right ]  \ , \\ 
M^* = \rho_0 e^{ - i\omega_0} \ \exp\left [ \frac{C_2}{4 N^2 \beta \rho_0^2 \sqrt{C_1 C_2}} \ \left (   G_0(m_1) - G_0(m_2) \right ) \right ]  \ , 
\end{eqnarray}
where $G_0(m)$ is the zero-distance Green function. 

Since we have two components of the mean Polyakov loop (
$\left\langle \TR U \right\rangle$ and 
$\left\langle \TR U^\dagger \right\rangle$, or equivalently
$\left\langle \RE \TR U \right\rangle$ and 
$\left\langle \IM \TR U^\dagger \right\rangle$), 
we have four components of the Polyakov loop correlation function, which can be gathered in the correlation matrix~\cite{bonati2018}
\begin{align}
	\label{corr-matrix}
	\Gamma(x,y) &= \begin{pmatrix}
		\left\langle \RE \TR U(x) \RE \TR U(y) \right\rangle & 
		\left\langle \RE \TR U(x) \IM \TR U(y) \right\rangle \\
		\left\langle \IM \TR U(x) \RE \TR U(y) \right\rangle & 
		\left\langle \IM \TR U(x) \IM \TR U(y) \right\rangle 
	\end{pmatrix}  \ \ . 
\end{align}
The off-diagonal terms vanish if $\mu = 0$, 
and the coefficients in the exponential decay of the connected parts in 
diagonal terms define the magnetic $m_M=m_1$ and electric $m_E=m_2$ screening masses. 
For $\mu > 0$ the electric and magnetic sector mix, so 
each correlation matrix element should be a sum of two terms
-- one decaying with $m_M$, and second with $m_E$. 
Eigenvalues of the correlation matrix~(\ref{corr-matrix}) turn out to be 
\begin{eqnarray}
{\cal{M}}_{1,2} = \frac{1}{2} \ M M^* \ 
\left [ F_1 \pm \sqrt{F_2(C_1) F_2(C_2)} \right ] \ , 
\label{eigenvalue1} 
\end{eqnarray}
\begin{equation}
F_1 \ = \ \exp\left [ \frac{1}{2 N^2 \beta \rho_0^2 } \ 
\left (   G_R(m_1) + G_R(m_2) \right ) \right ] - 1 \ , 
\label{F1_def}
\end{equation}
\begin{equation}
F_2(C_i) \ = \ \exp\left [ \frac{C_i}{2 N^2 \beta \rho_0^2 \sqrt{C_1 C_2}} \ \left (   G_R(m_1) - G_R(m_2) \right ) \right ] - 1 \ .
\label{F2_def}
\end{equation}
Notice that, if $m_1\leq m_2$, one obtains in the limit of large separation $R$: 
\begin{equation}
{\cal{M}}_{1} \ = \ \frac{M M^*}{2N^2\beta \rho_0^2} \ G_R(m_1) \ , 
\label{M1}
\end{equation} 
\begin{equation}
{\cal{M}}_{2} =  
\begin{cases}
\frac{M M^*}{2N^2\beta \rho_0^2} \ G_R(m_2) \ , \ 
\mbox{if} \ m_1 \leq m_2 \leq 2 m_1 \ ,  \\ 
\frac{M M^*}{2N^2\beta \rho_0^2} \left (  \ G_R(m_2) + \frac{1}{16 N^2\beta\rho_0^2} 
 \left ( 2-\frac{C_1+C_2}{\sqrt{C_1 C_2}} \right ) \ G_R^2(m_1) \right )  \ , \ 
\mbox{if} \ m_2 \geq 2 m_1 \ . 
\end{cases}
\label{M2}
\end{equation} 
Then, for example, the connected part of the Polyakov loop correlation becomes
\begin{eqnarray}
\label{PL_corr}
\langle \ {\rm Tr}U(0) {\rm Tr}U^*(R) \ \rangle  = M M^* (G_R(m_1) + G_R(m_2)) 
\sim {\cal{M}}_{1} + {\cal{M}}_{2} \ . 
\end{eqnarray}
The masses $m_1$ and $m_2$ play the role of the magnetic and electric screening masses and are given by 
\begin{eqnarray}
	\label{masse_def}
	m_{1,2} &=&  \frac{1}{2 \beta \rho_0^2} \ 
	\left ( C_3 \mp \sqrt{C_1 C_2}  \right ) \ , \\
	\label{C12_def}
	C_{1,2} &=& b_2 + (d \beta  - b_1) \rho_0^2 \pm i b_3 \rho_0 \ ,  \ 
	C_3 =  b_2 + (b_1 - d \beta ) \rho_0^2 \ . 
\end{eqnarray}
These expressions for screening masses are the main result of the mean-field analysis. Let us make a few comments on these expressions. 

\begin{itemize}
\item 
Using explicit expressions for the coefficients $b_i$, one proves that $m_2 \geq 2 m_1$, {\it i.e., the electric mass is at least two times larger than the magnetic mass}. Monte-Carlo simulations support this conclusion. 
\item 
If $\mu$ is non-vanishing then $b_3\ne 0$ and masses may, in principle, become complex and $m_1=m_2^*$. In this case the correlation would decay exponentially and exhibit oscillatory behavior as follows from Eq.~(\ref{PL_corr}). Such behavior is typical for a liquid phase which exists in $Z(3)$ spin model in the presence of external complex magnetic field~\cite{oscillating_phase} and in the large-$N$ limit of the $SU(N)$ Polyakov-loop model~\cite{LargeN_PLmodel_corr}. However, for the $SU(3)$ model studied here we were not able to find a set of parameters for which the product $C_1 C_2$ is negative and masses are complex. Either the oscillating region is very narrow or it is not realized in this model. 
\item
 The behavior versus $\beta$ of the two screening masses predicted by analytic expressions~(\ref{masse_def}),~(\ref{C12_def}) is shown
 in Fig.~\ref{fig:mass-gap-analytic}. This can be compared with results of 
 Monte-Carlo simulations presented below. One finds a reasonable qualitative, and sometime even quantitative, agreement. 
\end{itemize}

One can see that both masses grow when going away from the transition point, with the higher mass having very rapid growth after entering the ordered phase. 
Both masses have their minimum at the transition point, with lower mass getting closer to zero at the transition. 
\begin{figure}[H]
	\centering{
		\includegraphics[width=0.3\textwidth]{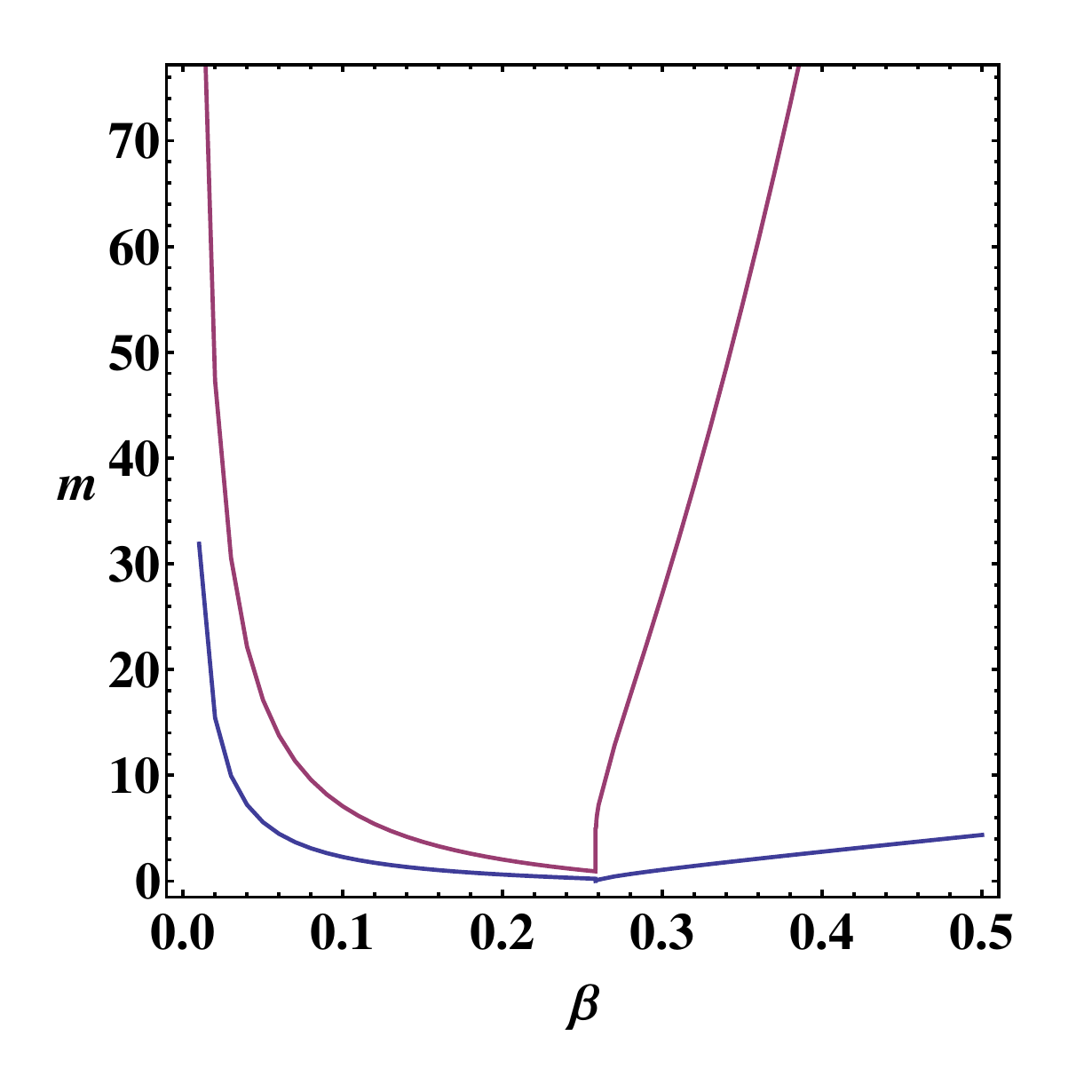}
		\includegraphics[width=0.3\textwidth]{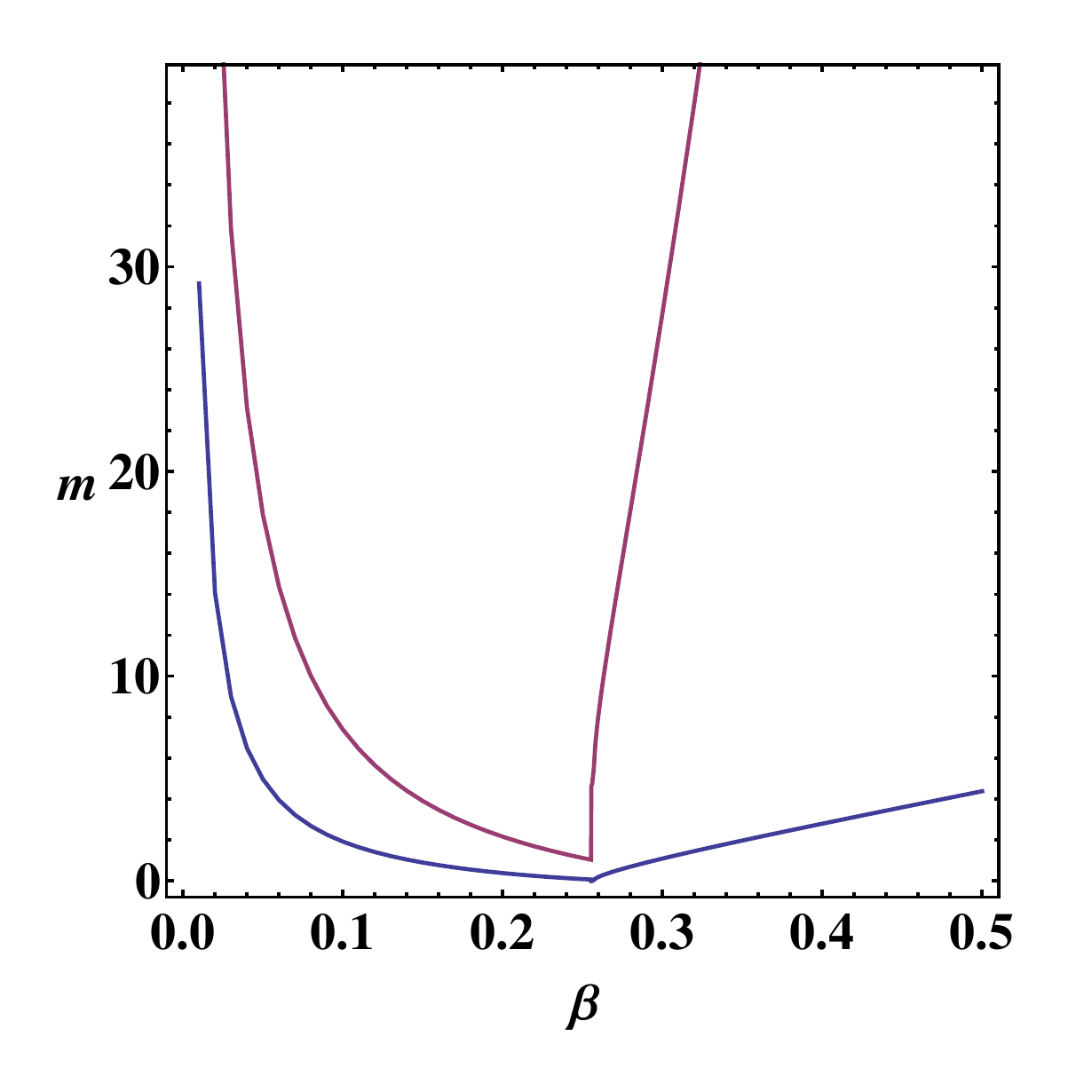} 
		\includegraphics[width=0.3\textwidth]{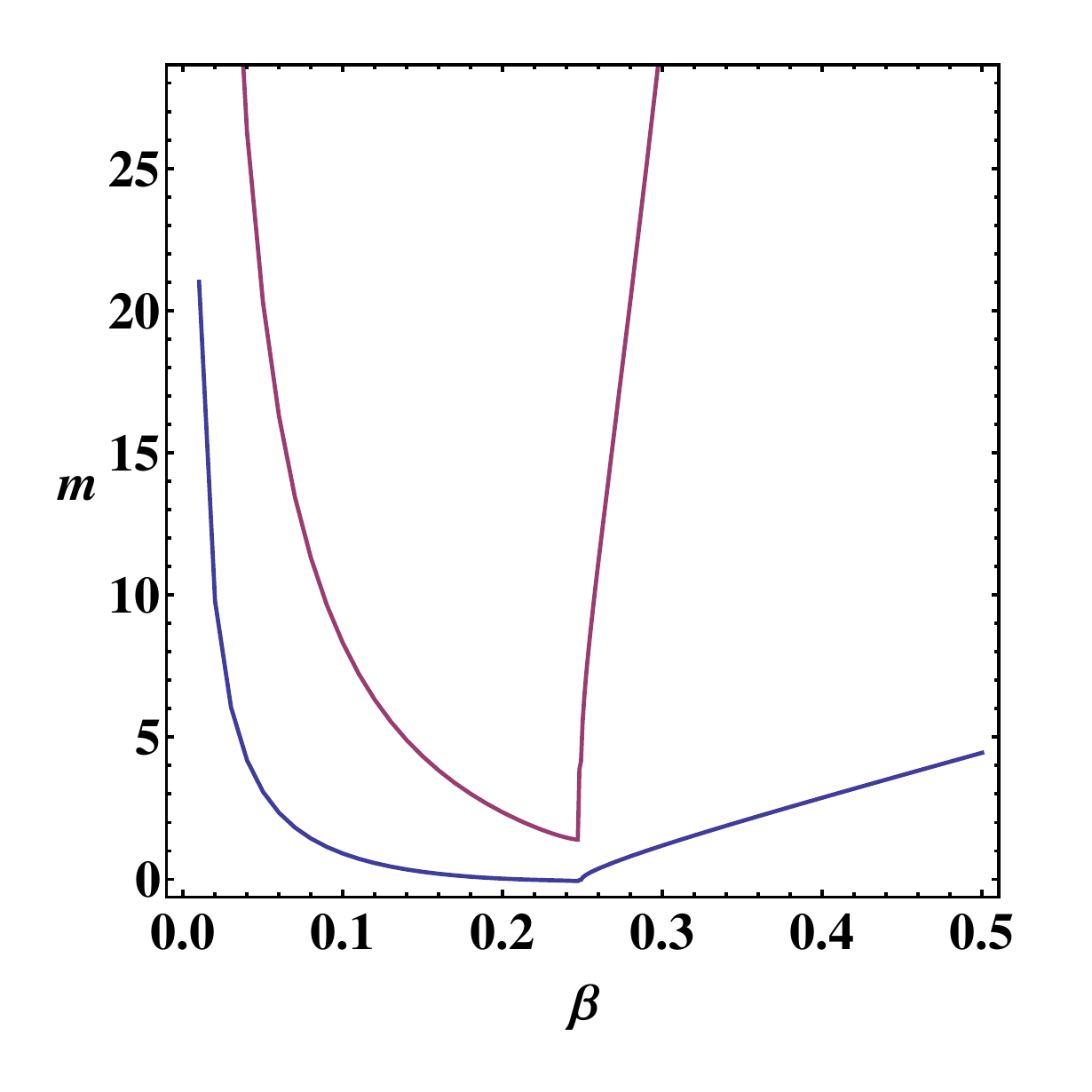}
	}
	\caption{Dependence on $\beta$ of the screening masses $m_1$ (blue curve) and $m_2$ (red curve) obtained from the analytic
          expressions~(\ref{masse_def}),~(\ref{C12_def}) 
		for $h=0.01$, $\mu=0$ (left), $h=0.01$, $\mu=0.9635$ (center), 
		and $h=0.01$, $\mu=2.0$ (right). }
	\label{fig:mass-gap-analytic}
\end{figure}

\section{Monte-Carlo results for correlations and screening masses} 

In this Section we present results of simulations and describe the behavior of the screening masses extracted from the exponential decay of the Polyakov-loop correlations. These masses will be compared with the masses computed from the second-moment correlation length. Details of the lattice setup and update are described in Ref.~\cite{Borisenko:2020cjx}. In this study we used lattices with sizes $L=20,24$ and sometimes $L=32$. The phase diagram described in Ref.~\cite{Borisenko:2020cjx} was obtained from simulations mainly on the lattice $L=20$. The larger lattices we used here gave us the possibility to quantitatively improve the phase diagram of the model. Such improved diagram is briefly described in the Appendix. 

\subsection{Two-point correlation functions}

To see the impact of non-zero chemical potential on the correlation function 
behavior, we calculated the values of the two-site correlation functions for 
several values of parameters. 
We considered six kinds of correlation functions:
\begin{align}
 \Gamma_{\mathrm{nn}}(R) &= \left\langle \Tr U(0) \Tr U(R) \right\rangle \ ,& 
 \Gamma_{\mathrm{rr}}(R) &= \left\langle \Re \Tr U(0) \Re \Tr U(R) \right\rangle \ ,
 \nonumber \\
 \Gamma_{\mathrm{nc}}(R) &= \left\langle \Tr U(0) \Tr U^\dagger(R) \right\rangle \ ,&
 \Gamma_{\mathrm{ri}}(R) &= \left\langle \Re \Tr U(0) \Im \Tr U(R) \right\rangle \ ,
 \\
 \Gamma_{\mathrm{cc}}(R) &= \left\langle \Tr U^\dagger(0) \Tr U^\dagger(R) \right\rangle \ ,&
 \Gamma_{\mathrm{ii}}(R) &= \left\langle \Im \Tr U(0) \Im \Tr U(R) \right\rangle \ .
 \nonumber
\end{align}
The correlation matrix~(\ref{corr-matrix}) takes the form
\begin{align}
	\label{corr-matrix_notation}
	\Gamma(R) = \begin{pmatrix}
		\Gamma_{\mathrm{rr}}(R) & 
		\Gamma_{\mathrm{ri}}(R) \\
		\Gamma_{\mathrm{ri}}(R) & 
		\Gamma_{\mathrm{ii}}(R) 
	\end{pmatrix} \ .  
\end{align} 

In the dual formulation the correlation functions can be written as
\begin{align}
 \Gamma_{\mathrm{nn}}(R) &= \left\langle \frac{R_3(n(0) + 1, p(0))}{R_3(n(0), p(0))} 
				\frac{R_3(n(R) + 1, p(R))}{R_3(n(R), p(R))} \right\rangle \ ,
 \nonumber \\
 \Gamma_{\mathrm{nc}}(R) &= \left\langle \frac{R_3(n(0) + 1, p(0))}{R_3(n(0), p(0))} 
				\frac{R_3(n(R), p(R) + 1)}{R_3(n(R), p(R))} \right\rangle \ ,
 \label{gamma_dual} \\
 \Gamma_{\mathrm{cc}}(R) &= \left\langle \frac{R_3(n(0), p(0) + 1)}{R_3(n(0), p(0))} 
				\frac{R_3(n(R), p(R) + 1)}{R_3(n(R), p(R))} \right\rangle \ .
 \nonumber
\end{align}
These formulas work for $R > 0$. For $R = 0$ both shifts to the $n$, $p$ variables 
happen at the same point, so only one ratio remains. The correlations $\Gamma_{\mathrm{rr}}$, 
$\Gamma_{\mathrm{ri}}$ and $\Gamma_{\mathrm{ii}}$ can be obtained as linear combinations of $\Gamma_{\mathrm{nn}}$,
$\Gamma_{\mathrm{nc}}$ and $\Gamma_{\mathrm{cc}}$.

The expressions~(\ref{gamma_dual}) become unusable when $h = 0$, and can 
have bad convergence properties for very small $h$, or very large $\mu$ values. 
We have checked by comparing the numerical results with the strong coupling expressions for small $\beta$ values, 
that the results can be relied on for $h > 0.005$ and $\mu < 3$. 

Since we work at non-zero $h$, the average traces can become non-zero, making the large-distance correlation function constant even in the disordered phase. Due to that, we introduce the connected correlation functions:
\begin{align}
 \Gamma_{\mathrm{nn}, \mathrm{conn}}(R) &= \Gamma_{\mathrm{nn}}(R) - \left\langle \Tr U \right\rangle^2 \ ,
 \nonumber \\
 \Gamma_{\mathrm{nc}, \mathrm{conn}}(R) &= \Gamma_{\mathrm{nc}}(R) - \left\langle \Tr U \right\rangle \left\langle \Tr U^\dagger \right\rangle \ ,
  \\
 \Gamma_{\mathrm{cc}, \mathrm{conn}}(R) &= \Gamma_{\mathrm{cc}}(R) - \left\langle \Tr U^\dagger \right\rangle^2 \ .
 \nonumber
\end{align}
One expects an exponential decay for these connected correlations
\begin{equation}
 \Gamma(R) = A \ \frac{\exp(-m R)}{R} \ ,
\label{corr-mass-gap}
\end{equation}
at least in the disordered phase.

\begin{figure}[H]
\centering{
  \includegraphics[width=0.31\textwidth]{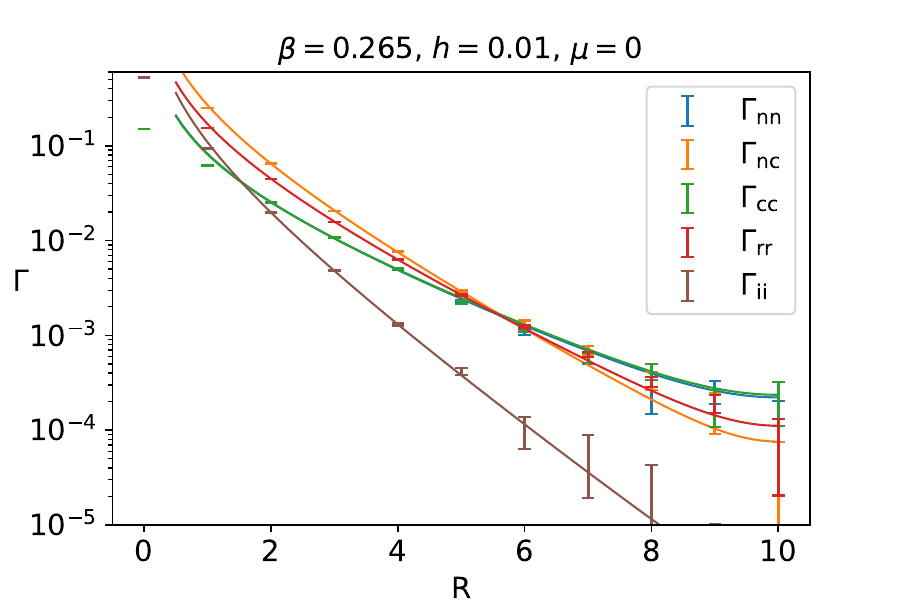}
  \includegraphics[width=0.31\textwidth]{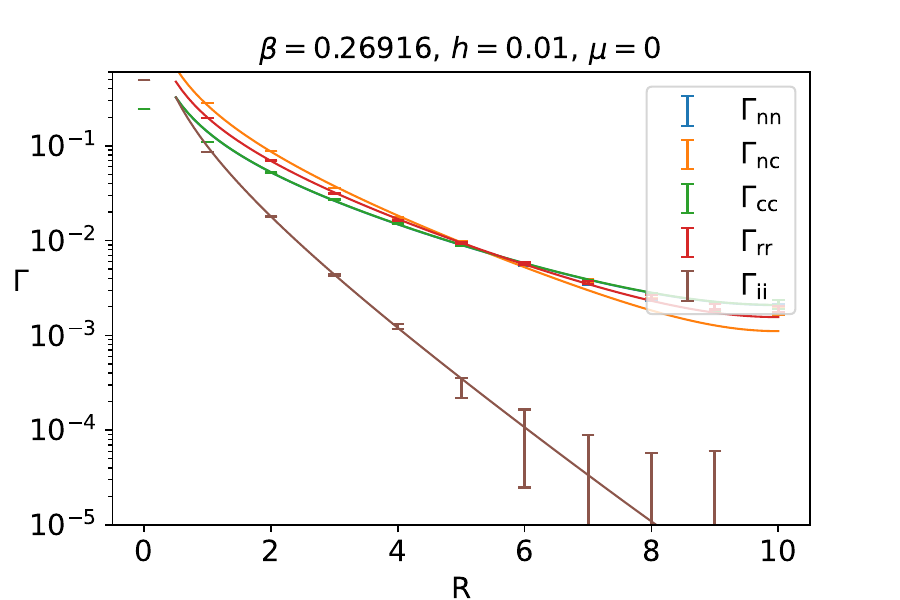} 
  \includegraphics[width=0.31\textwidth]{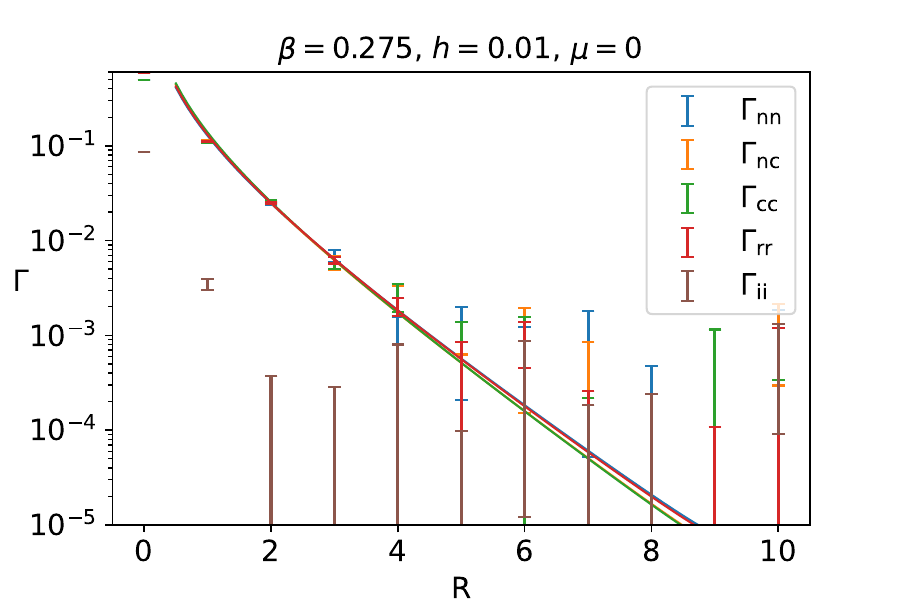} \\
  \includegraphics[width=0.31\textwidth]{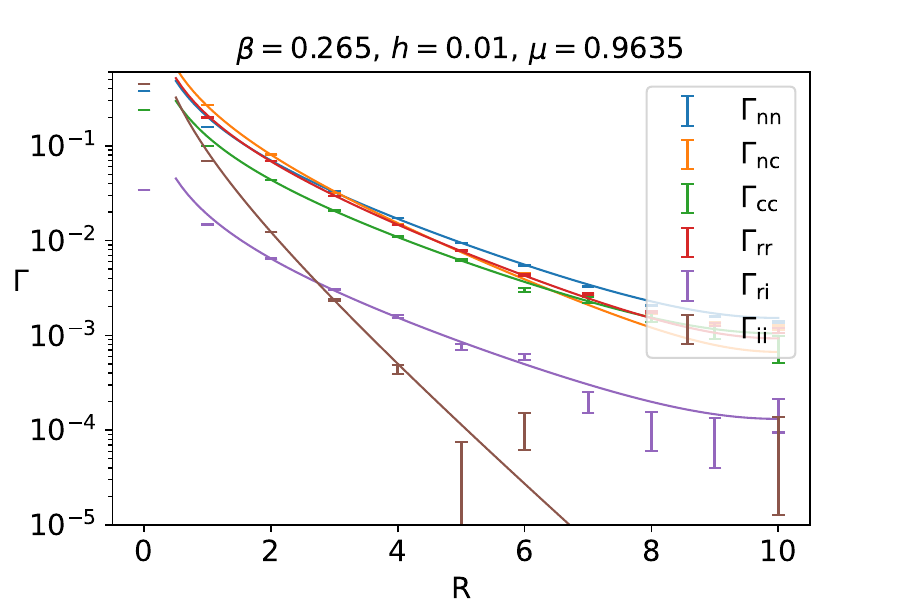} 
  \includegraphics[width=0.31\textwidth]{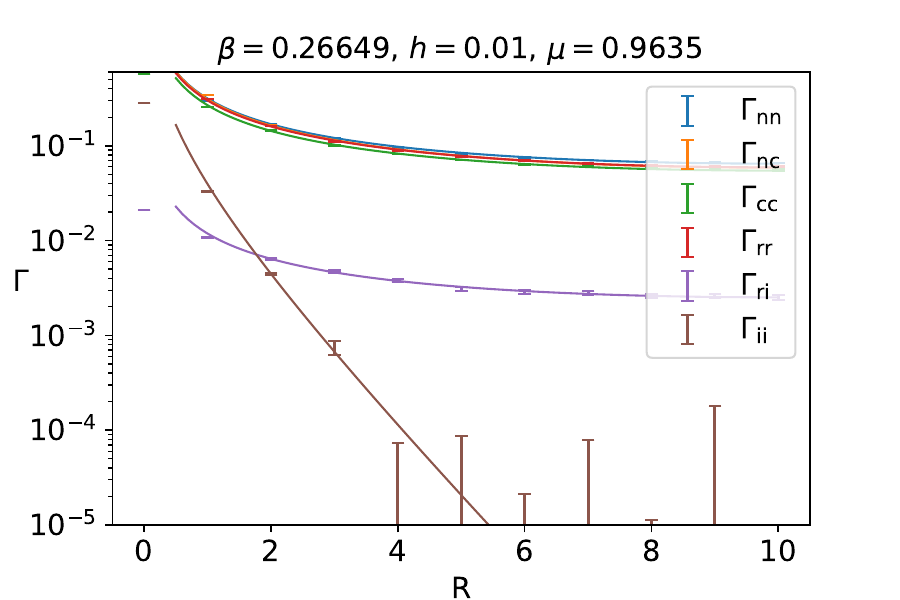}
  \includegraphics[width=0.31\textwidth]{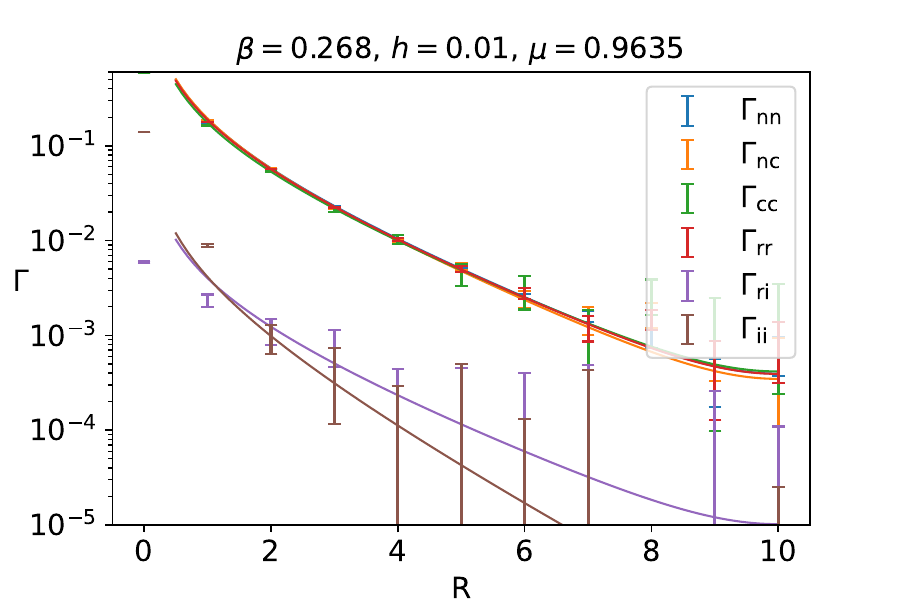} \\
  \includegraphics[width=0.31\textwidth]{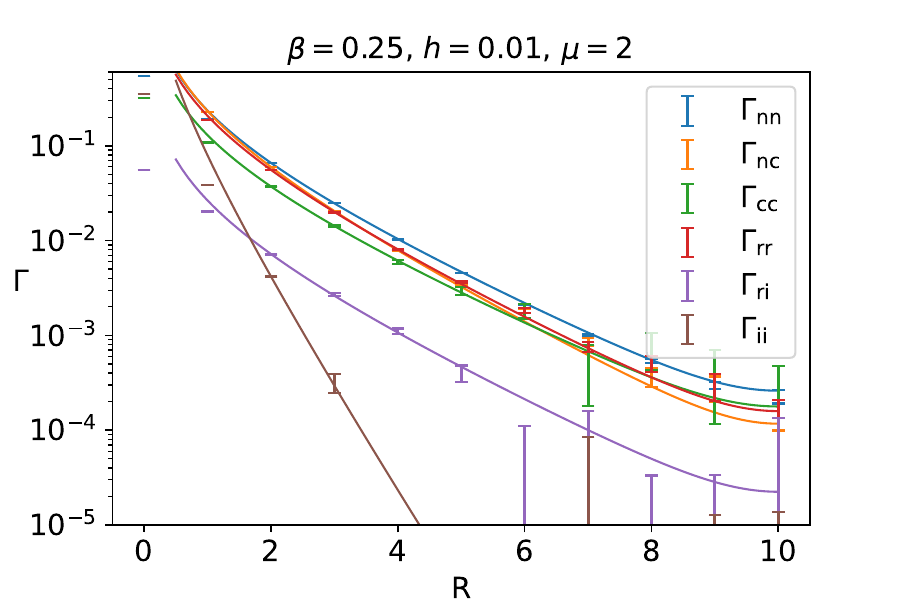}
  \includegraphics[width=0.31\textwidth]{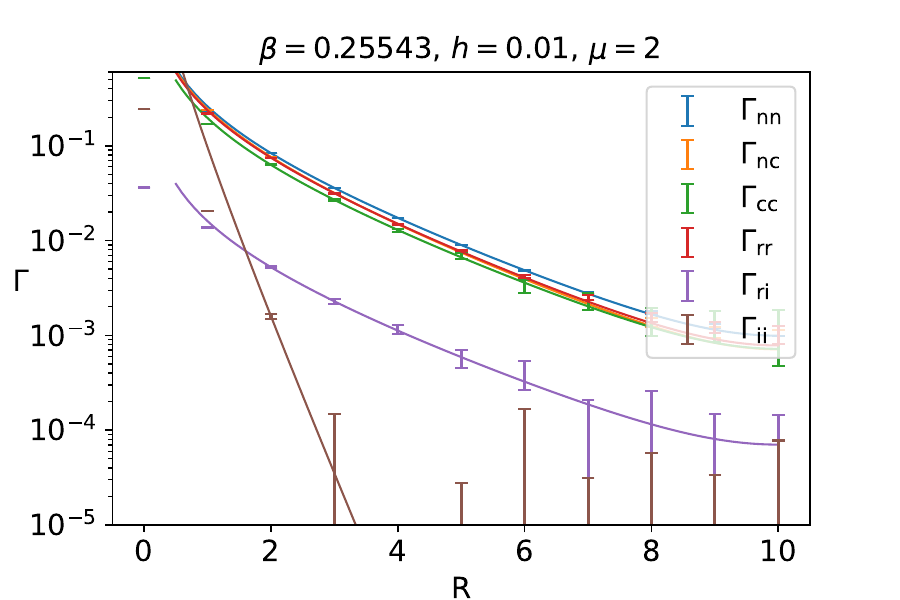}
  \includegraphics[width=0.31\textwidth]{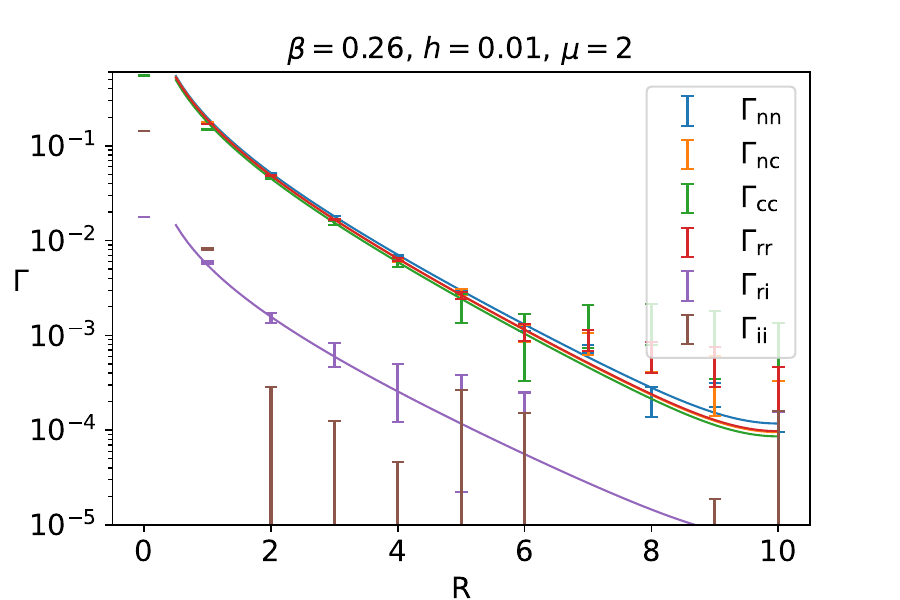} 
}
\caption{Behavior of the correlation functions on lattice with $L=20$ 
for different values of parameters $\beta$, $h$ and $\mu$. 
Solid lines represent the best fit to the function (\ref{corr-mass-gap}) with periodic contribution ($R \to L - R$).
Plots on the left are in the disordered phase, 
 in the middle -- near the phase transition (crossover) point,
 on the right -- in the ordered phase.}
\label{fig:corr}
\end{figure}

Samples of correlation function behavior at different points of the phase diagram are shown 
in Fig.~\ref{fig:corr}. One can see that, indeed, both in disordered and ordered phases,
the correlations decay exponentially. An interesting property is that, while the mass gap, 
corresponding to the slope of the plots, remains the same for $\Gamma_{\mathrm{nn}}$,
$\Gamma_{\mathrm{nc}}$, $\Gamma_{\mathrm{cc}}$, $\Gamma_{\mathrm{rr}}$ and $\Gamma_{\mathrm{ri}}$ correlation functions,
it is much larger for the $\Gamma_{\mathrm{ii}}$ correlation function. 
Also the mass gap for the $\Gamma_{\mathrm{ii}}$ correlation remains more or less constant, and in particular
does not vanish in the vicinity of the phase transition. 

At $\mu = 0$ the off-diagonal terms $\Gamma_{\mathrm{ri}}(R)$ are zero, and one can define 
magnetic and electric correlation masses as the exponential decay rates for, correspondingly, 
real-real and imaginary-imaginary connected correlations,
\begin{align}
	\label{em-screening-mu-zero}
	\Gamma_{\mathrm{rr}, \mathrm{conn}}(R) &\sim \frac{e^{-m_M R}}{R} \ , & \Gamma_{\mathrm{ii}, \mathrm{conn}}(R) &\sim \frac{e^{-m_E R}}{R} \ .
\end{align} 

At non-zero $\mu$ the correlation matrix has non-zero off-diagonal elements, and each element has contributions from two masses:
\begin{align}
	\label{em-screening-mu-nonzero}
	\Gamma_{\mathrm{conn}}(R) &=  A_1 \frac{e^{-m_1 R}}{R}  + A_2 \frac{e^{-m_2 R}}{R} \ .
\end{align}
While for small $\mu$ values the coefficients for the contribution of the smaller mass to $\Gamma_{\mathrm{ii}}$, 
and of the larger mass to $\Gamma_{\mathrm{rr}}$ are small, they make the extraction of the masses (especially the larger one) difficult.
Diagonalizing the correlation matrix $\Gamma(R)$, one can get contributions that depend purely on $m_1$ or $m_2$ like in $\mu=0$ case, 
which we fit to the exponential decay behavior to obtain the two masses. 

The masses extracted from fitting the diagonalized correlation matrix are shown in Fig.~\ref{fig:mass-gap}. 
One can see that the smaller mass behavior defines the transition: it rapidly drops to a small, but non-zero value, at the 
first-order transition point; it drops to zero (or, more precisely to a small value that goes to zero with $L \to \infty$) 
at the second-order transition point; it has a smooth minimum in the case of crossover.
The higher mass remains constant and of the order of one in the disordered phase, and starts to grow rapidly around the transition,
thus meaning that the corresponding correlation drops very fast in the ordered phase and vanishes at distances of a couple of lattice
spacings, making therefore very difficult to estimate the mass in that phase.  Such behavior agrees well with the mean-field prediction $m_2\geq 2 m_1$.

\subsection{Second moment correlation length}

Since in many cases the distances on which the correlation length can be determined reliably does not exceed five lattice spacings,
the mass gap extracted from fitting the exponential decay of the correlation functions has a large uncertainty.

An alternative to this approach is to extract from data second-moment correlation length of the (diagonalized) correlation functions $\Gamma_1$, $\Gamma_2$: 
\begin{align}
 \chi_j &= \frac{1}{L^3} \sum_R \Gamma_{j}(R) \ , \nonumber \\
 F_j &= \frac{1}{L^3} \sum_R \Gamma_{j}(R) \cos \frac{2 \pi R_1}{L} \ , 
 \label{xi2-def} \\
 \xi_{2, j} &= \frac{\sqrt{\frac{\chi_j}{F_j} - 1}}{2 \sin \frac{\pi}{L} } \ , \nonumber
\end{align}
where the summation is taken over all vectors $R = (R_1, R_2, R_3)$ on the lattice.

Assuming $\Gamma_{j}(R) = A \frac{e^{-m R}}{R}$, and taking the thermodynamic limit $L \to \infty$, 
we can replace summation by the integration, and trigonometric functions by their Taylor expansions.
Then the integration gives $\xi_{2,j} = \frac{1}{m} + O(1/L)$. 
Existence of larger masses $m_k$ in the correlation function spectrum introduces a finite correction that becomes larger 
when the ratio $m_k/m$ becomes closer to $1$. 
Thus, in general $\xi_{2,j} \neq \frac{1}{m}$ even in thermodynamic and continuum limit. 
We introduce  a ``second-moment mass gap'' $m^{(\xi_2)}_j = 1/\xi_{2,j}$ as an approximation to the actual mass gap $m$. 
To make this approximation closer to the actual mass gap, and to attempt extracting $m_2$, we 
have to perform the diagonalization of the correlation matrix and take the second moment of the diagonalized correlation functions, 
thus removing corrections from $m_2$ to $m^{(\xi_2)}_1$.
Since $\chi_j$ and $F_j$ are linear in $\Gamma_{j}$, we apply diagonalization to the $\chi$ and $F$ matrices to obtain
$\chi_{1,2}$ and $F_{1,2}$ -- the moments of the correlation functions corresponding to the lower and higher masses. 

\begin{figure}[hbt]
\centering{
  \includegraphics[width=0.327\textwidth]{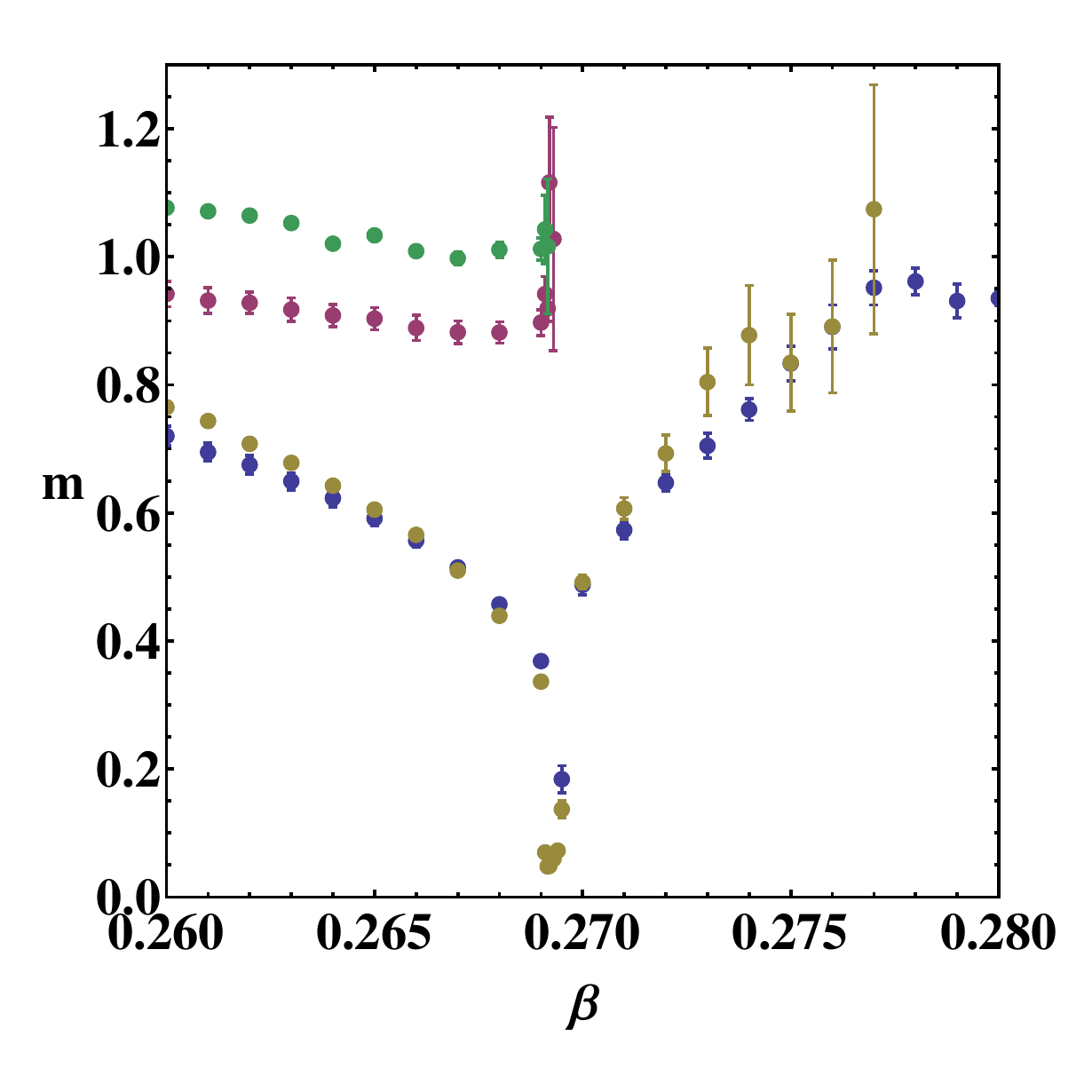}
  \includegraphics[width=0.327\textwidth]{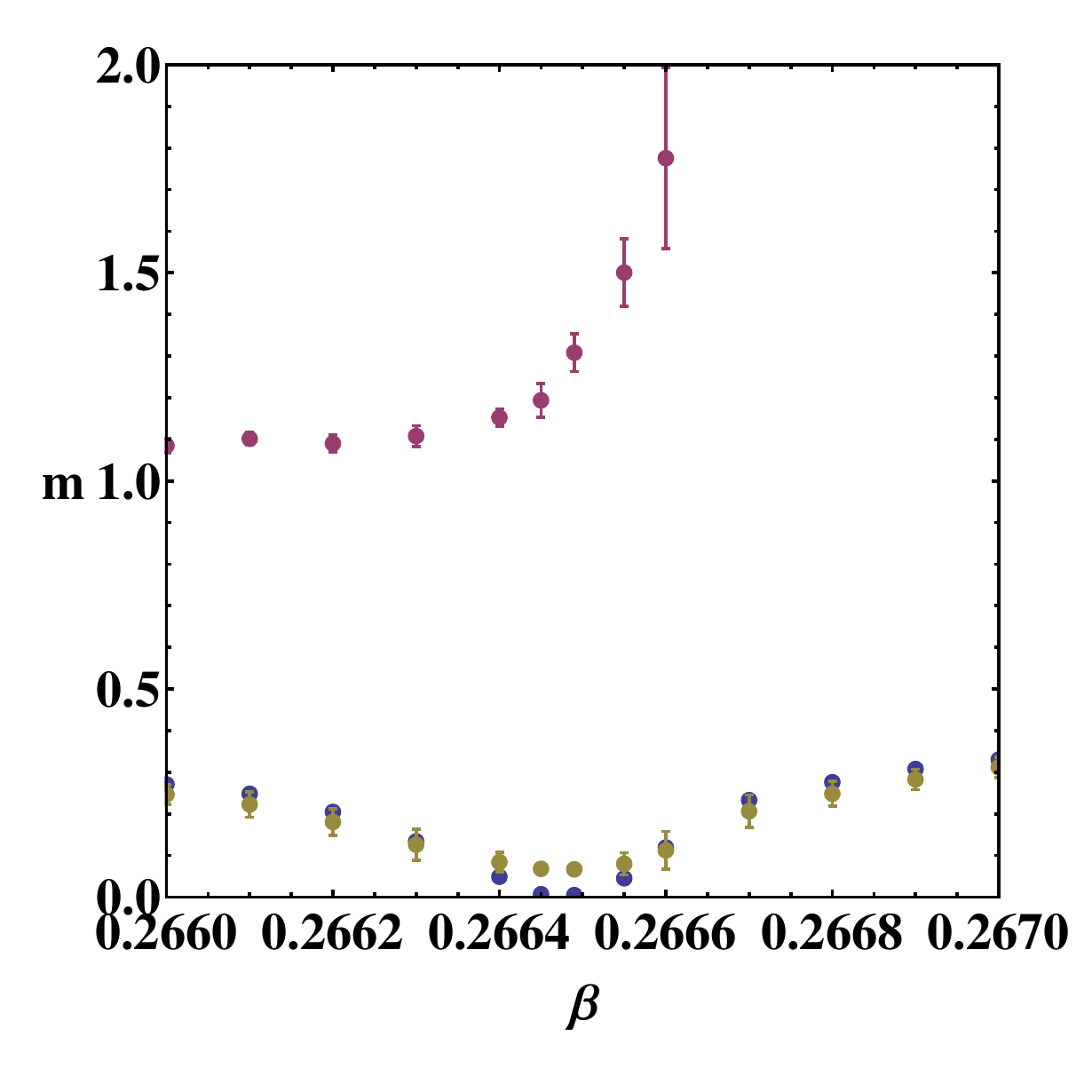} 
  \includegraphics[width=0.327\textwidth]{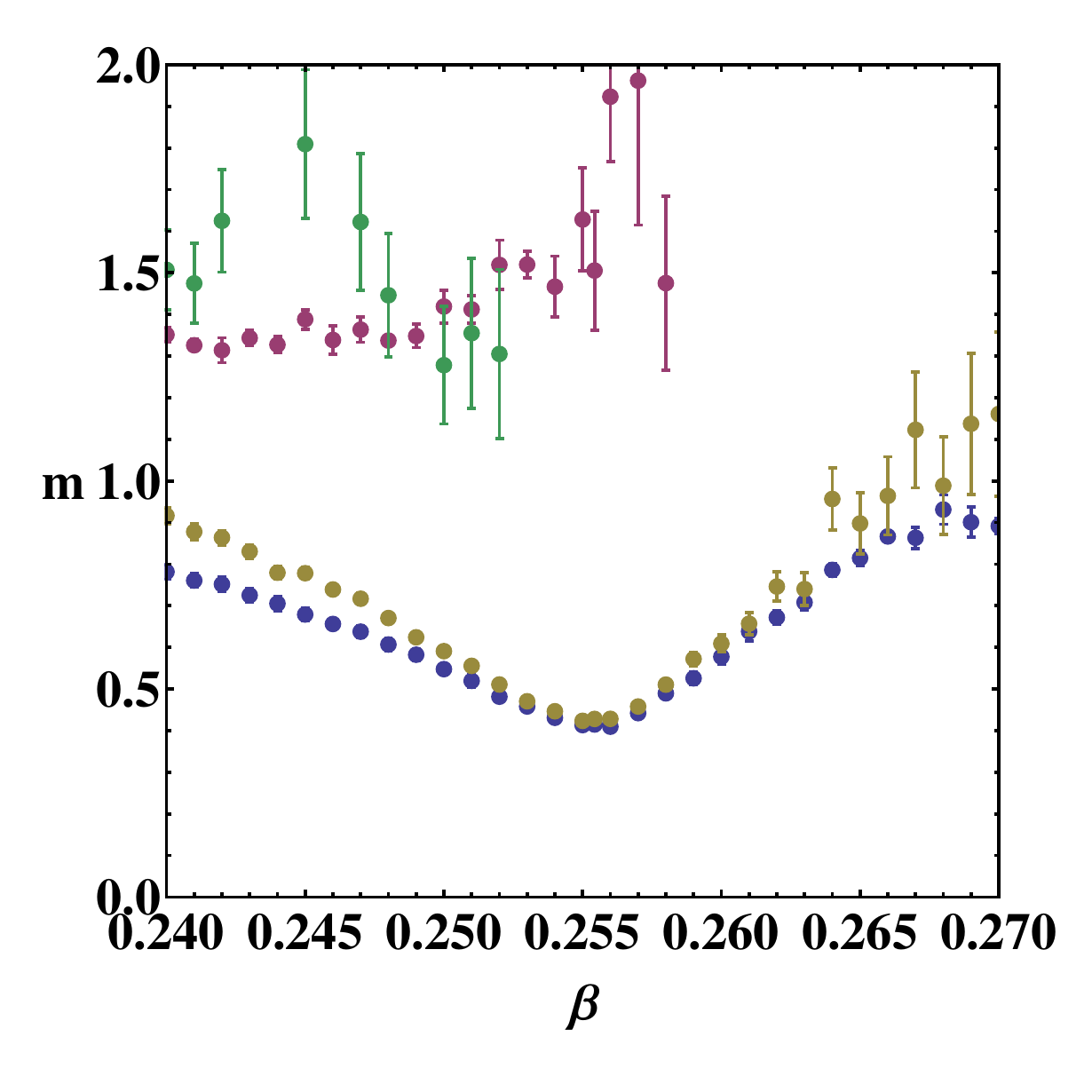}
}
\caption{Dependence on $\beta$ of the lower and higher screening masses extracted from the diagonalized correlation function fits (blue and violet markers), and 
from the diagonalized second-moment correlation lengths (yellow and green markers) around the transition point
for $h=0.01$, $\mu=0$, $L=16$ (left, first order phase transition), $h=0.01$, $\mu=0.9635$, $L=24$ (center, second order phase transition), 
and $h=0.01$, $\mu=2.0$, $L=16$ (right, crossover). The data points where the error estimate is above 1/3 of the mass value are removed for plot clarity.}
\label{fig:mass-gap}
\end{figure}

The ``second-moment mass gaps'' extracted in this way 
are compared with the mass gaps extracted from the correlation function fits in Fig.~\ref{fig:mass-gap}.
One can see that the lower masses extracted from the correlation function fits and from the second-moment correlation length
are in good agreement, especially close to the transition. The higher masses get very large errors in the ordered phase for $\mu=0$, 
and the diagonalization needed for non-zero $\mu$ results in large errors for the higher masses at any $\beta$. The behavior of the
higher mass in the disordered phase at $\mu = 0$ is the same as for the mass extracted from the correlation function fit, up to a shift, that can be explained by a corrections from higher masses.

 \section{Summary}

In this paper we studied the dual formulation of $SU(3)$ lattice gauge theory with one flavor of static staggered fermions. In this approximation the original theory can be first mapped onto an effective Polyakov-loop model. The latter can be further mapped onto a dual formulation with a positive weight.
The main idea of this study is to reveal the behavior of the screening masses and the second moment correlation length at finite baryon density. We used a simple mean-field approximation and Monte-Carlo simulations to accomplish such study.
Let us briefly summarize our main results.

\begin{itemize}

\item
Using the larger lattices and better statistics we have quantitatively improved the phase diagram presented in Ref.\cite{Borisenko:2020cjx}. No qualitative changes have been found.

\item
To get an idea of how the screening masses behave, we used the mean-field approximation to calculate the correlation matrix of the Polyakov loops.
Two screening masses have been extracted analytically from the correlations of the real and imaginary parts of the Polyakov loops. In all cases their behavior agrees with expectations. Especially good agreement with Monte-Carlo data is found in the vicinity of the second-order phase transition.

\item
We have computed numerically all possible correlations of the Polyakov loops.
The screening masses and the second-moment correlation length have been obtained from the fitting of the connected correlations to the exponential decay.
In all three cases (first-, second-order phase transitions and crossover) one observes a reasonable agreement with the expectations and with the mean-field results for the lower magnetic mass. The second, higher mass grows very rapidly after the transitions, preventing us from obtaining good fits. 
The general trend is that when $\mu$ is increased,
the screening masses exhibit a less steep variation across transition when the
coupling $\beta$ (which corresponds to the temperature in the underlying
QCD theory) is increased.

\item
Both the mean-field and numerical results do not show the existence of the complex mass spectrum in the theory which was found in two-dimensional QCD with the static quarks \cite{Ogilvie10,Ogilvie16}, in $Z(3)$ spin model with the complex magnetic field \cite{oscillating_phase} and in the large-$N$ limit of the :Polyakov-loop model similar to the one studied here \cite{LargeN_PLmodel_corr}.

\end{itemize}

The two main approximations used to obtain the dual formulation with a positive Boltzmann weight are the strong coupling approximation for the Wilson action and calculation of the quark determinant in the static approximation.
In Ref.\cite{Borisenko20} we have presented  a dual formulation valid at all values of the temporal gauge coupling constant. This formulation preserves the positivity of the Boltzmann weight. Therefore, one of the possible direction for the future work is an extension of the present simulations to the dual representation derived in \cite{Borisenko20}. Another direction is to include the leading corrections to the static quark determinant. It is not yet clear at the moment if such corrections preserve the positivity of the dual weight. This problem is currently under investigation.

\vspace{0.5cm}

{\bf \large Acknowledgments}

\vspace{0.2cm}

Numerical simulations have been performed on the ReCaS Data Center of INFN-Cosenza.
O.B. and A.P. acknowledge support from the INFN/NPQCD project.
V.C. acknowledges support by the Deutsche Forschungsgemeinschaft (DFG, German Research Foundation) through the 
CRC-TR 211 `Strong-interaction matter under extreme conditions' -- project number 315477589 -- TRR 211.
E.M. was supported in part by a York University Graduate Fellowship Doctoral - International.
This work is (partially) supported by ICSC – Centro Nazionale di Ricerca in High Performance Computing, Big Data and Quantum Computing,
funded by European Union – NextGenerationEU.

\vspace{0.5cm}

\appendix
\section{Improved phase diagram}\label{sec:derivemag}

The larger lattices used in the present work create the possibility to refine the phase diagram 
of the model in the $(h, \mu)$-plane. As in the previous work, we performed a standard finite-size scaling analysis on the
peak value of the magnetization susceptibility $\chi$.
Since we had to explore a three-parameter space, we could not afford to
perform high-statistic simulations if not on lattices with linear
sizes $L=16$ and 24. We determined $\chi$ for several $\beta$ values in the
transition region and fitted them to a Lorentzian, thus getting the position
of the peak, which gives the pseudocritical coupling $\beta_{\rm pc}$, and its
height. Comparing the dependence of the peak height on the lattice size $L$
with the scaling law
\begin{equation}
	\label{sp_Height_scaling}
	\chi_L(\beta_{\rm pc}) = A L^{\gamma/\nu}\;,
\end{equation}
we estimated the critical-index ratio $\gamma/\nu$ and collected all
our determinations, as many as 165, in 
Table~\ref{table:beta_critic_and_gamma_over_nu_versus_h_nu}.

We can see that, within uncertainties, the values of $\gamma/\nu$ are
spread in a range between 3, which implies a first-order transition, and
0, which holds for crossover, passing through the second order
3-dimensional Ising value, $\gamma/\nu=1.9638(8)$~\cite{PELISSETTO2002549}.
These sparse values of $\gamma/\nu$ are evidently an artifact of the relatively
small lattice sizes we could simulate. If we could approach the thermodynamic
limit, we would see that values of $\gamma/\nu$ concentrate around the values
of 3. (first order), 1.9638(8) (second order in the 3-dimensional Ising class)
and 0 (crossover). This is expected since we know that at $\mu=0$ in the
pure gauge limit of QCD, or for heavy enough quark masses, there is a whole
region of first-order deconfinement transitions in the $m_{\rm u,d}$-$m_s$ plane
(the famous Columbia plot), delimited by a line of second-order critical
points in the 3-dimensional Ising class~\cite{Karsch:2000xv}: thereafter, for
lower quark masses, the crossover region is met. In the simulations of our
effective Polyakov-loop model at non-zero density toward the thermodynamic
limit we should see the continuation of the line of second-order critical
points to non-zero values of the chemical potential. For the lattice volumes
considered in our study, we are not able to make a clear-cut
assignment of each choice of the parameters $h$ and $\mu$ to one of the three
transition regions. Using the determination of $\gamma/\nu$, we tried to
make this assignment, extending and modifying the three possible options
(first order, second order and crossover) as seen in
Table~\ref{table:gamma_over_nucolor}. This makes no sense in the
thermodynamic limit, but can be helpful in the present context.
In Table~\ref{table:gamma_over_nucolor} we introduced a color code, to
identify which of these regions a given parameter pair $(h,\mu)$ falls in.

\begin{table}[htb]
	\centering 
	\begin{tabular}{ | c | c | c |}
		\hline \hline
		$\gamma/ \nu$  					& color			   &phase\\
		\hline \hline
		$\gamma/\nu\geq $3 				&green			   &first order\\
		2.50 $\leq\gamma/\nu<$ 3		&light green	   &more first order than second order\\
		1.98 $<\gamma/\nu<$ 2.50		&yellow			   &more second order than first order\\
		1.94 $\leq\gamma/\nu\leq$ 1.98	&red			   &very close to second order\\
		1.85 $\leq\gamma/\nu<$ 1.94 	&brown			   &more second order than crossover\\
		0.3  $\leq\gamma/\nu<$ 1.85		&magenta		   &more crossover than second order\\
		0    $\leq\gamma/ \nu<$ 0.3		&blue			   &crossover\\
		\hline \hline
	\end{tabular}
	\caption{Color code used to characterize the different phases depending on 
		the value of $\gamma/\nu$. }
	\label{table:gamma_over_nucolor}
\end{table}

\begin{figure}[htb]
	\centering
	{\includegraphics[width=1.0\textwidth]{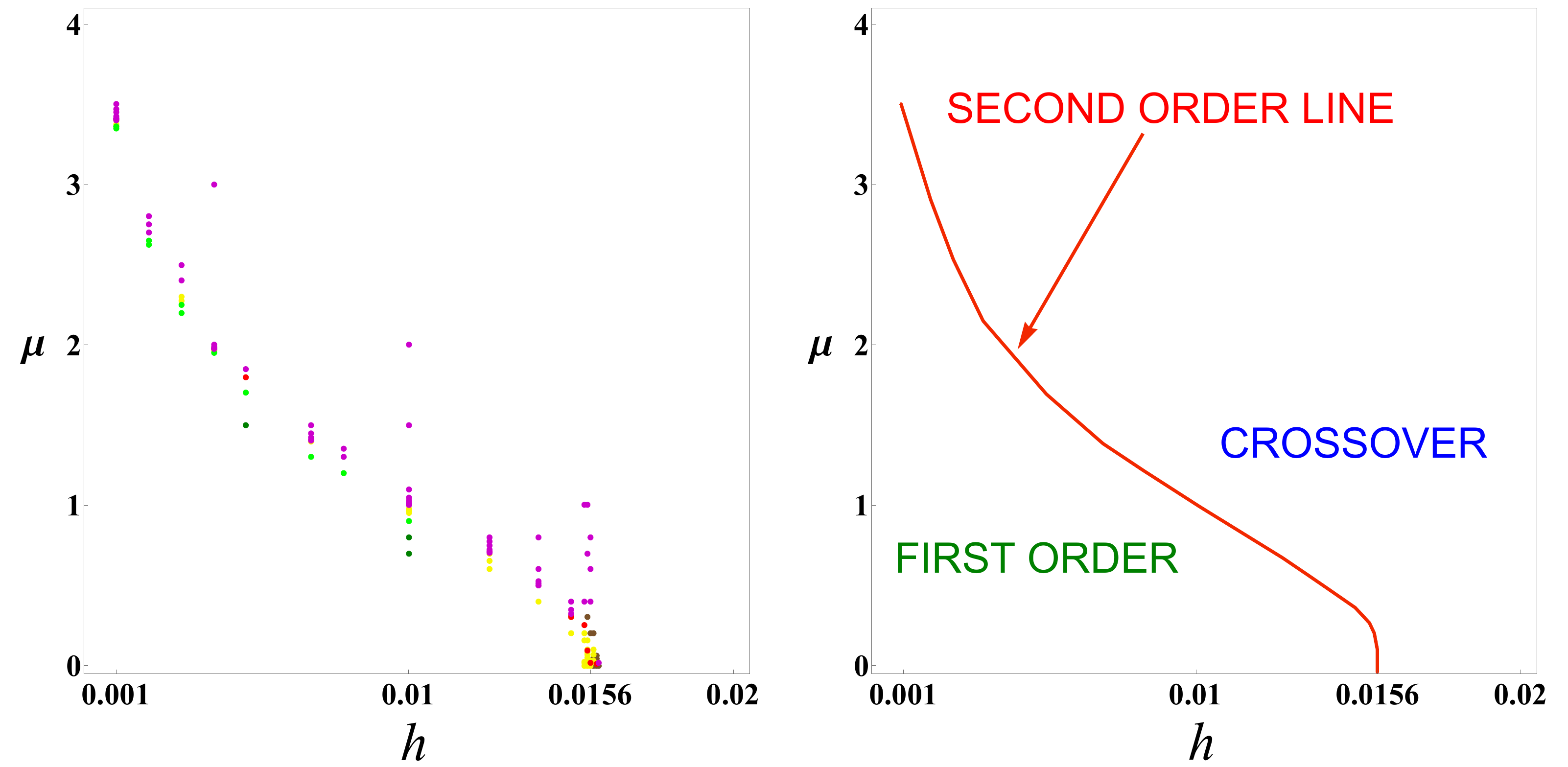}}
	\caption{(Left) Assignment of each parameter pair $(h,\mu)$ to a transition
		region according to the color code of Table~\ref{table:gamma_over_nucolor}.
		(Right) Estimated phase diagram.}
	\label{fig:Phase_diagram}
\end{figure}

In Fig.~\ref{fig:Phase_diagram}(left) each parameter pair $(h,\mu)$ considered
in our simulations is represented by a colored dot in the $(h,\mu)$ plane,
according to the color code defined in Table~\ref{table:gamma_over_nucolor},
allowing us to sketch a tentative phase diagram in the right panel
of the same figure.

We have not performed simulations in the absence of an external field. To find the
critical value $\beta_{\rm g}$ at $h=0$, {\it i.e.} for the pure gauge
theory, and at $\mu=0.$, we performed a simple fit of the form
$\beta_{\rm c}(h) = \beta_{\rm g}
+ a h$, as suggested in~\cite{Philipsen12}. We obtained the following values
$\beta_g=0.2741(2)$, $a=-0.50(2)$. The value of $\beta_{\rm g}$ agrees very well
with the value quoted in the literature, $\beta_{\rm g}=0.2741$ and reasonably
well with the mean-field result $\beta_{\rm g}=0.2615$. 

Another important question concerns the shape of the critical line shown in
the right panel of Fig.~\ref{fig:Phase_diagram} in the heavy-dense limit,
$h\to 0$, $\mu\to\infty$. From the data we have, one cannot make unambiguous
conclusions about its behavior. Nevertheless, data are well fitted by the
function $\mu_{\rm c}= - a \ln h - b h^2 + c $, with $a=0.988(17)$, $b=1406(201)$, $c=-3.4(1)$.

This shows that the line of second order phase transition might persist 
in the heavy-dense limit of QCD. 

\begin{landscape}
	\begin{table}
		\begin{tiny} 
			\begin{minipage}{0.6\textwidth}
				
				\begin{tabular}{| c | c | c | c |}
					\hline \hline
					$h$  		& $\mu$			   & $\beta_{\rm pc}$   & $\gamma/\nu$ \\
					\hline
					0.001&3.35&0.266777(2)&2.55(4)\\
					&3.36&0.266699(4)&2.52(6)\\
					&3.365&0.266658(1)&2.45(3)\\
					&3.375&0.266575(2)&2.45(5)\\
					&3.4&0.266378(2)&2.24(4)\\
					&3.403&0.266346(8)&1.64(6)\\
					&3.404&0.266349(1)&1.59(3)\\
					&3.405&0.266339(5)&1.56(4)\\
					&3.41&0.266307(2)&1.63(3)\\
					&3.415&0.266261(3)&1.52(4)\\
					&3.425&0.266179(10)&1.40(5)\\
					&3.45&0.265965(3)&1.18(5)\\
					&3.47&0.2658145(2)&1.10(3)\\
					&3.5&0.265537(1)&1.51(3)\\
					\hline
					0.002&2.625&0.266985(5)&2.68(11)\\
					&2.65&0.266800(2)&2.64(6)\\
					&2.7&0.266412(3)&1.69(5)\\
					&2.75&0.266001(3)&1.30(5)\\
					&2.8&0.265582(8)&1.50(12)\\
					\hline
					0.003&2.2&0.267080(3)&2.79(10)\\
					&2.25&0.266716(4)&2.58(9)\\
					&2.275&0.266522(1)&2.35(4)\\
					&2.3&0.266326(1)&2.14(3)\\
					&2.4&0.265502(7)&0.81(6)\\
					&2.5&0.264587(7)&1.06(4)\\
					\hline
					0.004&1.95&0.266752(1)&2.55(3)\\
					&1.96&0.266670(3)&2.42(6)\\
					&1.97&0.266599(5)&2.26(5)\\
					&1.975&0.266555(2)&2.34(5)\\
					&1.9755&0.266547(3)&1.83(6)\\
					&1.9756&0.266550(1)&1.78(4)\\
					&1.9757&0.266558(1)&1.76(4)\\
					&1.976&0.266549(2)&1.86(3)\\
					&1.978&0.266537(4)&1.86(5)\\
					&1.98&0.266548(3)&1.89(3)\\
					&1.99&0.266445(2)&1.71(3)\\
					&2&0.266363(3)&1.68(5)\\
					&3&0.25371(5)&0.76(5)\\
					\hline
					0.005&1.5&0.268127(1)&3.44(5)\\
					&1.7&0.266864(4)&2.68(9)\\
					&1.8&0.266110(2)&1.95(4)\\
					&1.85&0.265705(2)&1.64(3)\\
					\hline
					0.007&1.3&0.266394(3)&2.84(13)\\
					&1.4&0.266384(3)&2.21(8)\\
					&1.405&0.266359(1)&1.62(4)\\
					&1.41&0.266330(1)&1.63(2)\\
					&1.425&0.266211(2)&1.44(4)\\
					&1.45&0.266025(2)&1.42(5)\\
					&1.5&0.265639(4)&1.54(4)\\
					\hline
					0.008&1.2&0.266716(2)&2.56(4)\\
					&1.3&0.266023(4)&1.85(4)\\
					&1.35&0.265658(9)&1.57(4)\\
					\hline
					0.01&0.7&0.267824(3)&3.04(13)\\
					&0.8&0.267352(3)&3.17(18)\\
					&0.9&0.266837(2)&2.66(8)\\
					&0.95&0.266553(3)&2.40(8)\\
					&0.96&0.266491(3)&2.31(7)\\
					
					\hline	 \hline
				\end{tabular}
			\end{minipage}
			\begin{minipage}[c]{0.6\textwidth}
				\begin{tabular}{| c | c | c | c |}
					\hline \hline
					$h$  		& $\mu$			   & $\beta_{\rm pc}$   & $\gamma/\nu$ \\
					\hline	 
					0.01&0.963&0.266472(2)&2.26(4)\\
					&0.96325&0.266470(2)&2.32(4)\\
					&0.9635&0.266470(2)&2.41(6)\\
					&0.964&0.266472(2)&2.35(5)\\
					&0.96425&0.266465(2)&2.29(4)\\
					&0.9645&0.266464(3)&2.29(6)\\
					&0.965&0.266463(3)&2.32(4)\\
					&0.967&0.266454(1)&2.28(3)\\
					&0.97&0.266426(2)&2.36(5)\\
					&0.98&0.266369(1)&2.29(5)\\
					&1&0.266247(4)&2.19(8)\\
					&1.001&0.266235(3)&1.66(4)\\
					&1.0025&0.266230(3)&1.65(4)\\
					&1.003&0.266226(3)&1.64(5)\\
					&1.004&0.266202(3)&1.57(3)\\
					&1.005&0.266212(5)&1.65(4)\\
					&1.006&0.266197(1)&1.65(4)\\
					&1.007&0.266199(9)&1.65(5)\\
					&1.01&0.266185(2)&1.63(2)\\
					&1.015&0.266160(10)&1.53(6)\\
					&1.02&0.266114(3)&1.62(5)\\
					&1.025&0.266084(4)&1.52(4)\\
					&1.05&0.265908(2)&1.25(7)\\
					&1.1&0.265575(4)&1.58(6)\\
					&1.5&0.261997(5)&0.66(3)\\
					&2&0.254994(25)&0.72(4)\\
					\hline
					0.0125	&0.6&0.266633(2)&2.44(4)\\
					&0.65&0.266402(4)&2.13(6)\\
					&0.7&0.266152(3)&2.09(6)\\
					&0.7005&0.266156(4)&1.70(4)\\
					&0.701&0.266148(7)&1.69(7)\\
					&0.7025&0.266147(2)&1.70(3)\\
					&0.705&0.266132(2)&1.67(7)\\
					&0.708&0.266121(3)&1.74(6)\\
					&0.71&0.266107(3)&1.68(8)\\
					&0.712&0.266098(3)&1.65(5)\\
					&0.715&0.266082(2)&1.68(6)\\
					&0.725&0.266027(2)&1.56(7)\\
					&0.75&0.265885(2)&1.34(3)\\
					&0.775&0.265757(3)&1.30(3)\\
					&0.8&0.265603(2)&1.54(4)\\
					
					\hline
					0.014&0.4&0.266496(2)&2.45(7)\\
					&0.5&0.266146(2)&2.02(4)\\
					&0.5002&0.266142(2)&1.72(2)\\
					&0.5005&0.266143(4)&1.74(2)\\
					&0.501&0.266143(1)&1.79(3)\\
					&0.5025&0.266136(2)&1.78(4)\\
					&0.505&0.266124(2)&1.83(4)\\
					&0.51&0.266104(4)&1.71(5)\\
					&0.515&0.266090(4)&1.70(4)\\
					&0.525&0.266040(4)&1.64(6)\\
					&0.6&0.265710(3)&1.66(4)\\
					&0.8&0.264557(3)&1.04(4)\\
					&		&			&		\\
					&		&			&		\\
					&		&			&		\\
					&		&			&		\\

					\hline \hline
				\end{tabular}
			\end{minipage}
			\begin{minipage}[c]{0.6\textwidth}
				\begin{tabular}{| c | c | c | c |}
					\hline \hline
					$h$  		& $\mu$			   & $\beta_{\rm pc}$   & $\gamma/\nu$ \\
					\hline
					0.015&0.2&0.266440(2)&2.27(6)\\
					&0.3&0.266241(1)&2.11(3)\\
					&0.302&0.266227(3)&2.00(5)\\
					&0.305&0.266218(2)&1.96(4)\\
					&0.307&0.266216(2)&1.90(4)\\
					&0.31&0.266205(2)&1.89(3)\\
					&0.315&0.266197(4)&1.89(4)\\
					&0.325&0.266160(2)&1.80(4)\\
					&0.35&0.266099(7)&1.84(8)\\
					&0.4&0.265943(1)&1.83(4)\\
					\hline
					0.0154&0&0.266396(2)&2.15(7)\\
					&0.0154&0.266399(3)&2.22(8)\\
					&0.02&0.266398(5)&2.11(9)\\
					&0.154&0.266292(2)&2.05(7)\\
					&0.2&0.266227(3)&2.02(8)\\
					&0.25&0.266129(2)&1.96(5)\\
					&0.4&0.265715(5)&1.57(5)\\
					&1&0.261937(15)&0.64(3)\\
					\hline
					0.0155&0&0.266340(1)&2.05(4)\\
					&0.0155&0.266340(3)&2.15(11)\\
					&0.05&0.266338(2)&2.03(5)\\
					&0.07&0.266321(2)&1.99(4)\\
					&0.08&0.266334(8)&2.05(5)\\
					&0.085&0.266320(2)&2.16(5)\\
					&0.09&0.266310(2)&1.94(5)\\
					&0.1&0.266305(4)&2.12(9)\\
					&0.155&0.266241(2)&2.10(6)\\
					&0.3&0.265963(2)&1.86(5)\\
					&0.7&0.264221(12)&0.79(5)\\
					&1&0.261877(16)&0.60(3)\\
					\hline
					0.0156&0&0.266298(1)&2.01(4)\\
					&0.0156&0.266290(1)&1.97(2)\\
					&0.02&0.266290(2)&2.09(4)\\
					&0.025&0.266293(2)&2.10(5)\\
					&0.0275&0.266286(3)&2.03(3)\\
					&0.03&0.266294(3)&2.10(7)\\
					&0.2&0.266117(2)&1.90(5)\\
					&0.4&0.265598(2)&1.58(4)\\
					&0.6&0.264730(5)&1.03(4)\\
					&0.8&0.263473(5)&0.72(2)\\
					\hline
					0.0157&0&0.266245(5)&1.86(6)\\
					&0.0157&0.266244(2)&1.91(5)\\
					&0.02&0.266237(2)&1.98(5)\\
					&0.03&0.266240(4)&1.99(9)\\
					&0.05&0.266230(2)&1.91(5)\\
					&0.07&0.266218(3)&2.09(6)\\
					&0.1&0.266198(3)&2.04(6)\\
					&0.2&0.266071(5)&1.91(6)\\
					\hline
					0.0158&0&0.266191(1)&1.89(3)\\
					&0.01&0.266195(3)&1.97(5)\\
					&0.0158&0.266190(2)&1.93(6)\\
					&0.04&0.266181(3)&1.88(5)\\
					&0.06&0.266179(3)&1.90(5)\\
					\hline
					0.01585&0&0.266166(1)&1.91(5)\\
					&0.01585&0.266157(4)&1.79(6)\\
					&		&			&		\\
					&		&			&		\\
					\hline \hline
				\end{tabular}
			\end{minipage}

			\caption{Summary of the determinations of $\beta_{\rm pc}$ and $\gamma/\nu$
				for different values of $h$ and $\mu$.}
			\label{table:beta_critic_and_gamma_over_nu_versus_h_nu}
		\end{tiny}
	\end{table}
\end{landscape}

\end{document}